\documentclass[12pt]{article}
\usepackage[english]{babel}
\usepackage[dvips]{graphicx}
\usepackage{mathtext}
\usepackage{amsmath}
\usepackage{amssymb}

\begin{document}

\title{\bf Gravitational stability of dark energy \\ in galaxies and clusters of galaxies}
\author{Bohdan Novosyadlyj, Maksym Tsizh, Yurij Kulinich\\ \\
\textit{Astronomical observatory of Ivan Franko National University of Lviv,}\\
\textit{Kyryla i Methodia str., 8, Lviv, 79005, Ukraine}}
\maketitle

{\small
We analyze the behavior of the scalar field as dark energy of the Universe in a
static world of galaxies and clusters of galaxies. We find the analytical
solutions of evolution equations of the density and velocity perturbations
of dark matter and dark energy, which interact only gravitationally, along with
the perturbations of metric in a static world with background Minkowski
metric. Using them it was shown that quintessential and phantom dark
energy in the static world of galaxies and clusters of galaxies is
gravitationally stable and can only oscillate by the influence of self-gravity.
In the gravitational field of dark matter perturbations it is able to
condense monotonically, but the amplitude of density and velocity perturbations
on all scales remains small. It was illustrated also, that the "accretion" of
phantom dark energy in the region of dark matter overdensities causes
formation of dark energy underdensities - the regions with negative
amplitude of density perturbations of dark energy.

\medskip

PACS numbers: 95.36.+x, 98.80.-k}

\section*{Introduction}

The dark energy, which is responsible for the accelerated expansion of the
Universe, is one of the biggest mysteries of the Nature with uncertain prospects of
unraveling. Thanks to rapid growth of quality and quantity of the observational
data sensible to the dark energy properties there is no doubt in its existence,
but the spectrum of possible models is still too wide. New key
experiments are necessary for  establishing which model corresponds
to the dark energy of our Universe. Hitherto they were focused on cosmological
scales: luminosity - redshift relation for supernovae Ia, angular size - redshift
relation for fixed linear scales (baryon acoustic oscillations, X-ray gas
in clusters of galaxies), integral Sachs-Wolfe effect etc. Detailed description
of the models of dark energy and their testing with cosmological observations can
be found in the books
\cite{Amendola2010,Wolschin2010,Ruiz2010,Novosyadlyj2013m}.

Evolution, properties and dynamics of different models of dark energy in the non-stationary
homogeneous Universe with Friedmann-Robertson-Walker metric (FRW) and its
influence on the formation of the large scale structure of the Universe is
studied in details (see
\cite{Amendola2010,Wolschin2010,Ruiz2010,Novosyadlyj2013m}). However, today
little is known about what is going on with it in gravitational traps, i.e.
gravitationally bound systems, and what type and accuracy of measurements
should be done on galactic scales to constraint the dynamical dark energy. In the
papers of E. Babichev, V. Dokuchaev and Y. Eroshenko the analytical solutions
for the problem of accretion of dark energy on black holes with Schwarzschild
\cite{Babichev2004,Babichev2005}, Kerr-Newman \cite{Babichev2008}
and Reissner-Nordstr\"om \cite{Babichev2011} metrics are obtained and in the work
\cite{Babichev2012} the impact of such accretion on the space-time metric is
analyzed. Formation and structure of the halo of scale of clusters of galaxies
in the medium with dark matter and dark energy were studied with numerical
methods in \cite{Dutta2007,Abramo2007,Mota2008,Wang2009,Basse2011,Wang2012}. The
overall conclusion of these papers is that in galaxies and even in the vicinity
of black holes the density of the dark energy in the form of a scalar field,
which can either thicken or rarefy under the influence of gravitational
fields, is much less than the density of the gravitationally unstable dark
matter.

The goal of this paper is to analyze the gravitational instability of
dark energy on the background of dark matter inhomogeneities in the global
static Minkowski world based on common analytical solutions of the evolution
equations for linear perturbations of density, velocity and space-time metric.

\section{Average values of dark matter and dark energy \\ densities in
galaxies and clusters of galaxies}

We assume that dark energy is in the form of scalar field with the constant parameter of equation of state (EoS),
$w_{de}\equiv p_{de}/\rho_{de}c^2=const<-1/3$, which fills the Universe homogeneously in the early epoch before galaxies
and clusters formation. For estimation of amplitudes and characteristic scales of theory of the gravitational
instability we will evaluate the average densities of the components of the Universe in the clusters and galaxies for
values of cosmological parameters that fit the current data of observational cosmology:
\begin{equation}
\Omega_{de}=0.7,\quad  \Omega_{dm}=0.25,\quad \Omega_{b}=0.05,\quad H_0=70 \textrm{km/s/Mpc},
\label{cosm_par}
\end{equation}
and for EoS parameter we will use two values: $w_{de}=-0.9$ (quintessential scalar field) and $w_{de}=-1.1$
(phantom scalar field). $\Omega$'s denote the average values of density of dark energy ($de$), dark matter ($dm$) and
baryonic ($b$) components in units of critical density  $\rho_{cr}^{(0)}=3H_0^2/8\pi G$ in current epoch and $H_0$ is
Hubble parameter. These values of parameters are inside of 1$\sigma$ confidence intervals obtained from almost all
observational
data (see chapter 1 in \cite{Novosyadlyj2013m}). Henceforward we will use the value of parameter of matter density
$\Omega_m=\Omega_{dm}+\Omega_{b}=0.3$, since the dark matter and the baryonic matter on scales we are interested in are
described well by hydrodynamical approximation with EoS parameter of dust-like matter $w_{b}=w_{dm}=0$. In the current
epoch with  values of parameters (\ref{cosm_par}) the average densities of matter and dark energy are following:
\begin{equation}
 \rho_{m}^{u}=2.8\cdot10^{-30} \textrm{g/sm}^3, \quad \rho_{de}^{u}=6.4\cdot10^{-30} \textrm{g/sm}^3.
\label{rho_u}
\end{equation}

In clusters and galaxies the values of mean densities of matter and dark energy are, certainly, different
from these ones. Let us estimate them using the halo model of galaxies and clusters formation
\cite{Press1974,Smith2003,Kulinich2008,Sergijenko2009,Kulinich2012}. According to this model, the average
density of matter after virialization and establishing of the dynamical equilibrium (the virial theorem is valid) is
\begin{equation}
 \rho_{m}^{vir}=\Delta_{vc}\rho_{cr}(z_{col}),
 \label{rho_vir}
\end{equation}
where $\Delta_{vc}\approx100$ in current epoch, $\approx150$ at $z=1$ and $\approx180$ at $z>10$, and critical density
is taken for the moment of collapse of the central part of the uniform halo. For further estimation we assume that the
redshift of the collapse $z_{col}$ is 20 for typical galaxy and 1 for typical cluster of galaxies.
The critical density at current epoch in the model with the above parameters is
$\rho_{cr}^{(0)}=9.2\cdot10^{-30} \textrm{g/cm}^3$ and at any redshift it is
\begin{equation}
\rho_{cr}(z)=\rho_{cr}^{(0)}\left(\Omega_m(1+z)^3+\Omega_{de}(1+z)^{3(1+w_{de})}\right).  \nonumber
\end{equation}
Thus, for $z=1$ we get  $\rho_{cr}(1)\approx 3\cdot10^{-29} \textrm{g/cm}^3$ and for $z=20$
$\rho_{cr}(20)\approx 2.6\cdot10^{-26} \textrm{g/cm}^3$.
Taking into account (\ref{rho_vir}) we get average matter density in typical massive galaxy\footnote{Therefore average
dark matter density in the typical massive galaxy is
$\rho_{dm}^{gal}\approx 4\cdot10^{-24} \textrm{g/cm}^3$, which is 4 orders less then the value that may be detected in
the most precise measurements in Solar system \cite{Pitjev2013}. So there is no ground for "closing of hypothesis of
dark matter" due to upper limit given in \cite{Pitjev2013}: $\rho_{dm}^{gal}<1.1\cdot10^{-20} \textrm{g/cm}^3$.}
and rich cluster of galaxies
\begin{equation}
\rho_{m}^{gal}\approx 5\cdot10^{-24} \textrm{g/cm}^3, \quad  \rho_{m}^{cl}\approx 5\cdot10^{-27} \textrm{g/cm}^3.
\label{rho_m}
\end{equation}
Scalar field as dark energy, which almost uniformly fills the entire Universe, practically does not participate in the
virialization of the dark matter, but it feels the changes of the gravitational potential of the halo in the time and
space.

After halo (galaxy or cluster of galaxies) has been separated from the expansion of the Universe, which happens at the
moment
$z_{ta}$ ($t_{ta}=t_{col}/2$ in Einstein-de Sitter model), the dynamics of dark energy changes according to the dynamics
of local world. When halo after virialization is stabilized, the static world in its volume can be considered as
Minkowski world (we ignore here local inhomogeneities and curvature of space-time). For estimation we assume that the
dynamics of dark energy in galaxies and clusters stabilizes at their $z_{col}$ and from this moment its density does not
change. As
$\rho_{de}(z)=\Omega_{de}\rho_{cr}^{(0)}(1+z)^{3(1+w_{de})}$, for quintessential scalar field with $w_{de}=-0.9$ for
galaxy and cluster of galaxies we get
\begin{equation}
\rho_{de}^{gal}\approx 2\cdot10^{-29} \textrm{g/cm}^3, \quad  \rho_{de}^{cl}\approx 8\cdot10^{-30} \textrm{g/cm}^3,
\label{rho_de_q}
\end{equation}
and for phantom field respectively
\begin{equation}
\rho_{de}^{gal}\approx 2.5\cdot10^{-30} \textrm{g/cm}^3, \quad  \rho_{de}^{cl}\approx 5.2\cdot10^{-30} \textrm{g/cm}^3.
\label{rho_de_ph}
\end{equation}

\section{Perfect fluid small perturbation equations \\ in static Minkowski world}

We assume that static world with Minkowski metric is filled with weakly perturbed perfect fluid with stress-energy
tensor
\begin{equation}
T_{ij}=(c^2\rho+p)u_iu_j-g_{ij}p,
\label{tei}
\end{equation}
where $\rho(x^i)=\overline{\rho}+\delta\rho(x^i)$ is density of perfect fluid (in g/cm$^3$),
$p(x^i)=\overline{p}+\delta p(x^i)$ is its pressure (in erg/cm$^3$),
$u_i(x^i)=0+\delta u^i(x^i)$ is covariant 4-velocity $u^i=dx^i/ds$ (dimensionless values) and
$g_{ij}(x^i)=\gamma_{ij}+h_{ij}(x^i)$ is the metric tensor. The last one is the sum of the components of the Minkowski
metric tensor $\gamma_{ij}$ and small deviations, related to small curvature of space-time caused by perturbations of
density and velocity of energetic components of the Universe. Smallness of the perturbations infer, that
$\delta\rho(x^i)/\overline{\rho}\ll 1$,
$\delta p(x^i)/\overline{p}\ll 1$, $\delta u^i(x^i)\ll 1$, $|h_{ii}|\ll 1$, and give us the possibility to linearize the
equations - to drop the terms with quadratic and higher orders of perturbations. In case of scalar perturbations we can
write components of the metric as:
\begin{equation}
 h_{00}(x^i)=2\Psi(x^i), \quad h_{\alpha\alpha}(x^i)=-2\Phi(x^i), \quad  h_{0\alpha}(x^i)=0, \quad
 h_{\alpha\beta}(x^i)=0,\,({\alpha\ne\beta}), \nonumber
\end{equation}
which was shown in \cite{Bardeen1980} for FRW world. Thus, space-time metric is:
\begin{equation}
ds^2=(1+2\Psi(x^i))d\tau^2-(1+2\Phi(x^i))\delta_{\alpha\beta}dx^{\alpha}dx^{\beta},
\label{ds2}
\end{equation}
where $\tau\equiv ct$.
The frame in which metric can be reduced to the form (\ref {ds2}) is called conformal Newtonian frame and in the case
of FRW space it is comoving to the unperturbed cosmological background that expands according to the Hubble law. In
our case, the background (unperturbed) space-time is static, Hubble expansion is missing. If the fluid is isotropic
(anisotropic component of the stress tensor is equal to zero), then $\Psi = - \Phi$ is the actual potential of Newtonian
gravity.
Equation of the evolution of the relative perturbations of the density
$\delta(x^i)\equiv\delta\rho(x^i)/\overline{\rho}$ and
velocity $V^{\alpha}(x^i)\equiv\delta u^{\alpha}(x^i)$ can be obtained from differential equations of the conservation of the energy and
momentum $T_{i\,;j}^j=0$. After linearizing and Fourier decomposition
\begin{eqnarray}
\delta(x^{\alpha},\tau)&\equiv&\frac{1}{(2\pi)^3}\int e^{i\mathbf{k}\mathbf{x}}\delta_k(\tau)d^3k, \quad
\mathbf{V}(x^{\alpha},\tau)\equiv\frac{1}{(2\pi)^3}\int e^{i\mathbf{k}\mathbf{x}}\mathbf{V}_k(\tau)d^3k, \nonumber\\
\Psi(x^{\alpha},\tau)&\equiv&\frac{1}{(2\pi)^3}\int e^{i\mathbf{k}\mathbf{x}}\Psi_k(\tau)d^3k,\nonumber
\end{eqnarray}
we get the equations for Fourier modes of the relative perturbations of density and velocity
of the fluid:
\begin{eqnarray}
    \dot{\delta}_k + (1+w)kV_k-3(1+w)\dot{\Psi}_k  =   0,  \label{delta_k} \\
    \dot{V}_k- \frac{c_s^2k}{1+w}\delta_k-k\Psi_k = 0,\label{V_k}
\end{eqnarray}
where $w\equiv \overline{p}/\overline{\rho}c^2$ is EoS parameter of perfect fluid and $c_s^2\equiv \delta p/\delta\rho c^2$ is the effective sound speed (speed of propagation of perturbations in units of speed of light). Hereafter instead of vector of Fourier mode of velocity $\mathbf{V}_k(\tau)$ (or, in component representation $V^{\alpha}_k$) we will use its projection on the wave vector $\mathbf{k}$: $V_k\equiv \mathbf{k}\mathbf{V}_k/k$, where $k$ is vector modul of $\mathbf{k}$.

The Einstein equations give the equation for Fourier amplitude $\Psi_k$:
\begin{equation}
  \dot{\Psi}_k -\frac{4\pi G}{c^2}\rho(1+w)\frac{V_k}{k}=0.
  \label{Psi_k}
\end{equation}
System of equations (\ref{delta_k})-(\ref{Psi_k}) has a solution for given $w$ and $c_s^2$. Let us consider three cases
for which the analytical solutions exist.

Henceforth $(\dot{ })\equiv d/d\tau$ and it has dimension of cm$^{-1}$. All the Fourier amplitudes are dimensionless as
well as $w$ and $c_s^2$.

\section{Gravitational instability of dark matter}

For dark matter we take $w=c_s^2=0$ and equations (\ref{delta_k})-(\ref{Psi_k}) become
\begin{eqnarray}
    \dot{\delta}_k + kV_k-3\dot{\Psi}_k  =   0,  \label{delta1} \\
    \dot{V}_k - k\Psi_k = 0,\label{V1}\\
    \dot{\Psi}_k -\frac{4\pi G\rho}{c^2} \frac{V_k}{k}=0. \label{Psi1}
\end{eqnarray}
The two last equations are the first-order equations for two unknown functions and they can be rewritten as one equation
of the second order for one unknown function:
\begin{equation}
\ddot{\Psi}_k - \frac{4\pi G\rho}{c^2} \Psi_k =0, \label{Psi1a}
\end{equation}
which has the exact general analytical solution:
\begin{equation}
\Psi_k=C_1e^{\tau/\tau_m} + C_2e^{-\tau/\tau_m}, \label{Psi1m}
\end{equation}
which is superposition of two partial solutions: exponentially increasing and exponentially decreasing.
Here
$\tau_m\equiv c/\sqrt{4\pi G\rho_m}$ and it has the same dimension as $\tau$ does (i.e. cm). From the first two
equations (\ref{delta1})-(\ref{Psi1}) we get the general solutions for $\delta_k$ and $V_k$:
\begin{eqnarray}
\delta_k&=&(3-\tau_m^2k^2)\left(C_1e^{\tau/\tau_m} + C_2e^{-\tau/\tau_m}\right), \label{delta1m}\\
V_k&=&\tau_mk\left(C_1e^{\tau/\tau_m} - C_2e^{-\tau/\tau_m}\right), \label{V1m}
\end{eqnarray}

Let us estimate $\tau_m$ for medium with cosmological and galaxy average density of dark matter and also for average
density in clusters.

If we assume that matter density in the Universe is equal to the critical one, than
$$\tau_m=\sqrt{\frac{2}{3}}\,\frac{c}{H_0},$$
where [$H_0$]=sec$^{-1}$ and in model with (\ref{cosm_par}) $\tau_m\approx 3500$ Mpc or  $\approx 1\times 10^{10}$
years.
If the matter density $\rho_m=\Omega_m\rho_{cr}$, then $\tau_m$ must be multiplied by $\Omega_m^{-1/2}\approx 1.8$.
Therefore, $\tau_m$ is the time, during which (if the density is $\rho_m$) the amplitudes of perturbations of the
density, velocity and gravitational potential are increased in $e$ times. With average density in galaxy and in cluster
of galaxies (\ref{rho_m}) we get
\begin{equation}
\tau_m^{gal}\approx 4.7\, \textrm{Mpc} \quad (1.5\cdot10^7\,\textrm{years}) \quad \textrm{and} \quad
\tau_m^{cl}\approx 150\, \textrm{Mpc} \quad  (5\cdot10^8\,\textrm{years})
\label{tau_m_all}
\end{equation}
respectively.

Integration constant $C_1$ and $C_2$ in (\ref{Psi1m})-(\ref{V1m}) are determined from the initial conditions for every $k$. As we can see, sign of
the density $\delta_k$ is the same as $\Psi_k$ if $k<k_0\equiv \sqrt{3}/\tau_m$, and is opposite if $k>k_0$. Wavenumber
$k_0$ corresponds to scale $\lambda_0=2\pi/k_0\approx$23200 Mpc for $\rho_m^u$, 540 Mpc for $\rho_m^{cl}$ and 17 Mpc for
$\rho_m^{gal}$. This scale $k_0$ appears in relativistic approach of description of the perturbations evolution and
reflects the gauge dependence of the density perturbations \cite{Bardeen1980}. Indeed, if we neglect time dependence of
gravitational potential and replace equation (\ref{Psi1}) with Poisson equation than we immediately obtain
$\delta_k=-\tau_m^2k^2\Psi_k$, which arises from (\ref{delta1m}) for $\lambda\ll\lambda_0$. For structures with
$\lambda\gg\lambda_0$ we have $\delta_k=3\Psi_k$.

The scales of structures in galaxies are much smaller than $\lambda_0^{gal}$, the scales of galaxies and their groups in
clusters are much smaller than $\lambda_0^{cl}$ and scales of known structures in the Universe are much less
then $\lambda_0^{u}$. Thus, for known structures the signs of the perturbations of density and the gravitational
potential are opposite in both modes -- increasing and decreasing. Sign of $V_k$ is always the same as of $\Psi_k$ in
the increasing mode and is always opposite in the decreasing one.

For cosmology the most interesting is the growing mode of perturbation. If at initial moment of time $\tau=0$ there is
positive density perturbation $\delta_k=(3-\tau_m^2k^2)C>0$ ($\lambda\ll\lambda_0$) with negative potential $\Psi_k=C<0$
and negative velocity perturbation $V_k=\tau_mkC<0$ (motion towards the center of perturbation), then amplitudes of all
perturbed values will increase with exponential law\footnote{Result is known from classical work by J.H. Jeans
\cite{Jeans1902}.}.

\section{Gravitational stability of scalar field \\ as dark energy}

We assume that in this case, the same as in previous, the unperturbed quantities of density and pressure are constant
and velocities are zero. For dark energy in the form of the scalar field with $w_{de}<-1/3$ and $0\le c_s^2\le 1$  the
equations for perturbations are the complete system of three simple differential equations of the first order for three
unknown functions:
\begin{eqnarray}
    \dot{\delta}_k + (1+w_{de})kV_k-3(1+w_{de})\dot{\Psi}_k  =   0,  \label{delta2} \\
    \dot{V}_k- \frac{c_s^2k}{1+w_{de}}\delta_k-k\Psi_k = 0,\label{V2} \\
     \dot{\Psi}_k - \tau_{de}^{-2}(1+w_{de})\frac{V_k}{k}=0, \label{Psi2}
\end{eqnarray}
where $\tau_{de}=c/\sqrt{4\pi G\rho_{de}}$. In static Minkowski world with value of the dark energy density for current
epoch $\tau_{de}^u\approx 4200$ Mpc ($1.4\cdot10^{10}$ years).
This parameter for quintessential field ($w_{de}=-0.9$) in the model with cluster of galaxies density is
$\tau_{de}^{cl}\approx 3800$ Mpc ($1.2\cdot10^{10}$ years) and with galaxy density it is
$\tau_{de}^{gal}\approx 2400$ Mpc ($8\cdot10^{9}$ years). For phantom field ($w_{de}=-1.1$) values of these parameters
are also close: $\tau_{de}^{cl}\approx 4700$ Mpc ($1.5\cdot10^{10}$ years), $\tau_{de}^{gal}\approx 6700$ Mpc
($2\cdot10^{10}$ years). Equations (\ref{delta2})-(\ref{Psi2}) have analytical solutions if  $w_{de}$ and $c_s^2$ are
constant. Indeed, in this case equations (\ref{Psi2}) and (\ref{delta2}) imply:
\begin{equation}
    \delta_k= \left[3(1+w_{de})-\tau_{de}^2k^2\right]\Psi_k   \label{delta3}
\end{equation}
(integration constant is omitted, because it vanishes together with the perturbations).
If we now differentiate (\ref{Psi2}) with respect to $\tau$ and use (\ref{V2}) and (\ref{delta3}), then we obtain  the
equation
\begin{equation}
\ddot{\Psi}_k - \left[\frac{(1+3c_s^2)(1+w_{de})}{\tau_{de}^2}-c_s^2k^2\right] \Psi_k =0, \label{Psi2a}
\end{equation}
the type of solution of which depends on the sign of the expression in brackets: if it is positive, we have
exponential solutions, if it is negative -- the oscillating ones. We denote it as $\mathcal{D}^2$
and general solution of the system is rewritten as:
\begin{eqnarray}
\Psi_k&=&C_1e^{\sqrt{\mathcal{D}^2}\tau} + C_2e^{-\sqrt{\mathcal{D}^2}\tau}, \label{Psi2de}\\
\delta_k&=&\left(3(1+w_{de})-\tau_{de}^2k^2\right)\Psi_k,\label{delta_de2}\\
V_k&=&\frac{\tau_{de}^2k}{1+w_{de}}\sqrt{\mathcal{D}^2}\left(C_1e^{\sqrt{\mathcal{D}^2}\tau} -
C_2e^{-\sqrt{\mathcal{D}^2}\tau}\right).\label{V2de}
\end{eqnarray}

Let us analyze these solutions and find the condition of gravitational instability of such dark energy.

The general solution is unstable (has exponentially increasing mode) if $\mathcal{D}^2> 0$, this is the case for scales:
\begin{equation}
k< \frac{\sqrt{(1+3c_s^2)(1+w_{de})}}{c_s\tau_{de}}. \label{kj}
\end{equation}
This condition can be satisfied only for quintessential dark energy ($w_{de}>-1$) with positive square of the effective
sound speed.
The value $k_J^{de}\equiv \sqrt{(1+3c_s^2)(1+w_{de})}/(c_s\tau_{de})$ we will call wavenumber of Jeans scale for dark
energy. This scale in the Universe with $\rho_{de}=\Omega_{de}\rho_{cr}$ and $c_s^2=1$ is
\begin{equation}
\lambda_J^{de}=\frac{\pi\tau_{de}}{\sqrt{1+w_{de}}}, \nonumber
 \end{equation}
which gives  $\lambda_J^{de}\approx 10\tau_{de}$ for quintessence field with $w_{de}=-0.9$, which is much more than the
known structures in the Universe. Condition (\ref{kj}) is never satisfied for phantom fields. Thus, dark energy phantom
scalar fields are absolutely gravitationally stable.

Therefore, at the scales of galaxies and clusters both quintessential and phantom scalar fields are
gravitationally stable: they can only oscillate with a period $|\mathcal{D}|^{-1}$, determined by the scale of the
perturbation, speed of sound and the value of the average field density.

Note, that the sign of the density perturbations of quintessential scalar fields is opposite to the sign of gravitational potential similar to dark matter case if
\begin{equation}
k>k_0^{de}\equiv\frac{\sqrt{3(1+w_{de})}}{\tau_{de}}.
 \end{equation}
This corresponds to the scales of $\lambda<\lambda_0^{de}\approx11.5\tau_{de}$ Mpc. Interesting for cosmology and
astrophysics scales are in that range. Sign of velocity for quintessence coincides with the sign of the potential
for the growing mode and is opposite for the decaying one.

So, if we have a potential well ($\Psi_{de}<0$) with $\lambda>\lambda_0^{de}$, then quintessential dark energy flows
into it (velocity towards the center), the amplitude of the perturbation density increases, but sign of it, according
to (\ref{delta_de2}), is negative.
Obviously, this is consequence of relativism or gauge dependence of the density perturbations on the
super-large scales, where perturbations of density and metric are comparable: $\delta_{de}\approx3(1+w_{de})\Psi_{de}$.
On the medium scales $\lambda_J^{de}<\lambda<\lambda_0^{de}$ we have flowing of the dark energy inside the
"condensation" region, $\delta_{de}>0$. For perturbation scales $\lambda<\lambda_J^{de}$ dark energy oscillates
similarly to acoustic oscillation of baryon-photon plasma in pre-recombination epoch (see analytical solutions in
\cite{Novosyadlyj2007}).
For scales $\lambda\ll \lambda_J^{de} \quad (k\gg k_J^{de})$ real parts of the solutions (\ref{Psi2de})-(\ref{V2de})
with redefined constants have the form:
\begin{eqnarray}
\Psi_{de}&=&C_1\sin{(c_sk\tau)} + C_2\cos{(c_sk\tau)} , \label{Psi2de_d}\\
\delta_{de}&=&-\tau_{de}^2k^2\left[C_1\sin{(c_sk\tau)} + C_2\cos{(c_sk\tau)}\right],\label{delta2de_d}\\
V_{de}&=&\frac{c_s\tau_{de}^2k^2}{1+w_{de}}\left[C_1\cos{(c_sk\tau)} - C_2\sin{(c_sk\tau)}\right].\label{V2de_d}
\end{eqnarray}
The velocity oscillations in the case of quintessence take place with shifted by $+\pi/2$ phase and in the case of
phantom field they are shifted by $-\pi/2$.

\section{Two-component medium: dark energy \\ in the dark matter perturbations field}

Let us consider the Minkowski world, which is filled with two almost uniformly distributed components -- matter and dark
energy, which interact only gravitationally through the local perturbations of density and velocity. As before we assume
that both components -- matter and dark energy -- are described by the stress-energy tensor of perfect fluid
$T^{(m)}_{ij}$
and $T^{(de)}_{ij}$ respectively and are perturbed weakly in the sense described in section 2. In this approach the
equations for velocity and density perturbations are obtained from differential energy-momentum conservation law for
each component separately, $T^{j\,(m)}_{i\,;j}=0$,  $T^{j\,(de)}_{i\,;j}=0$, and equations for gravitational
field perturbations are the Einstein equations in metric (\ref{ds2}). In this case we have the system of five ordinary
differential equations for five unknown functions:
\begin{eqnarray}
\dot{\delta}_{m} + kV_{m}-3\dot{\Psi}_k  =   0,  \label{delta3m} \\
    \dot{V}_{m} - k\Psi_k = 0,\label{V3m}\\
     \dot{\delta}_{de} + (1+w_{de})kV_{de}-3(1+w_{de})\dot{\Psi}_k  =   0,  \label{delta3de} \\
    \dot{V}_{de}- \frac{c_s^2k}{1+w_{de}}\delta_{de}-k\Psi_k = 0,\label{V3de} \\
     \dot{\Psi}_k - \frac{4\pi G\rho_m}{c^2}\frac{V_{m}}{k} - \frac{4\pi G\rho_{de}}{c^2}(1+w_{de})\frac{V_{de}}{k}=0.
\label{Psi3}
\end{eqnarray}
Here $\delta_{m}$, $V_{m}$, $\delta_{de}$ and $V_{de}$ are Fourier amplitudes of the perturbations of density
and velocity of matter and dark energy respectively. In the galaxies and rich clusters of galaxies that can be
considered as a quasistatic world with the metric (\ref{ds2}) the average matter density is much higher than the
average density of dark energy ($\rho_m\gg\rho_{de}$ as it is evaluated in the section 1), that is why the dark energy
practically has no effect on the
gravitational instability of the matter, so for $\delta_m$, $V_m$ and $\Psi_k$ we can use solutions
(\ref{Psi1m})-(\ref{V1m}). Let us analyze the
behavior of the dark energy perturbations (equations (\ref{delta3de})-(\ref{V3de})) in the field of potential
(\ref{Psi1m}).

According to (\ref{Psi1a}) $\ddot{\Psi}_k=\Psi_k/\tau_m^2$,
so the equations (\ref{delta3de}) and (\ref{V3de}) are equivalent to inhomogeneous second-order equation for
$\delta_{de}$:
\begin{equation}
\ddot{\delta}_{de} + c_s^2k^2\delta_{de}=(1+w_{de})\frac{3-\tau_m^2k^2}{\tau_m^2}\Psi_k, \label{delta4de} \\
\end{equation}
which is the equation for forced oscillations. The differentiated equation (\ref{V3de}) together with (\ref{delta3de})
gives
inhomogeneous differential equation of the second order for $V_{de}$:
\begin{equation}
\ddot{V}_{de} + c_s^2k^2V_{de}=(1+3c_s^2)k\dot{\Psi}_k, \label{V4de} \\
\end{equation}
which is also equation for forced oscillations.

Equations (\ref{delta4de}) and (\ref{V4de}) have analytical solutions, which can be obtained by the method of variation
of constants
and which for the growing mode of dark matter perturbations $\Psi_k=A_k e^{\tau/\tau_m}$ (here $A_k$ - initial
amplitude of gravitational potential of dark matter) are as follows:
\begin{eqnarray}
\delta_{de}&=&\left[\tilde{C}_1\sin{(c_sk\tau)}+\tilde{C}_2\cos{(c_sk\tau)}\right]+
A_k\frac{(1+w_{de})(3-\tau_m^2k^2)}{1+c_s^2k^2\tau_m^2}e^{\tau/\tau_m}, \label{delta5de}\\
V_{de}&=& -\frac{c_s}{1+w_{de}}\left[\tilde{C}_1\cos{(c_sk\tau)}-\tilde{C}_2\sin{(c_sk\tau)}\right] +
A_k\frac{(1+3c_s^2)\tau_mk}{1+c_s^2k^2\tau_m^2}e^{\tau/\tau_m}, \label{V5de}
\end{eqnarray}
where $\tilde{C}_1$, $\tilde{C}_2$ are the constants of integration of homogeneous equation (\ref{delta4de}), which are
given by initial
conditions. In particular, they can be put equal to zero. For positive matter density
perturbation $\delta_{m}>0$ with scale
$\lambda<\lambda_0$ the gravitational potential is negative. The velocity amplitude of dark energy has the same sign,
which
means that it is flowing inside the region of perturbations of dark matter. Since $|A_k|\gg |C_1|,\,|C_2|$, the general
solution (\ref{delta5de})-(\ref{V5de}) describes monotonic flowing of dark energy into the gravitational well of the
dark matter (``accretion'') with low amplitude oscillations of density and velocity. For search of possible traces of
the influence of dark energy on structure formation of galaxies and clusters of galaxies the last term in
the solutions, describing the forced monotonic growth of the dark energy density and velocity perturbations amplitude
in the regions of high concentration of the matter, is important.

Let us compare the values of the density and velocity perturbations of dark energy and dark matter, taking the
appropriate ratio from (\ref{delta5de})-(\ref{V5de}) and (\ref{delta1m})-(\ref{V1m}),
\begin{equation}
\delta_{de}/\delta_m = \frac{(1+w_{de})}{1+c_s^2k^2\tau_m^2}, \quad V_{de}/V_m =\frac{(1+3c_s^2)}{1+c_s^2k^2\tau_m^2}.
\end{equation}
Values of ratio depend on characteristic time of dark matter gravitational instability $\tau_m$, scale of perturbation
$k$, effective sound speed of dark energy $c_s$ and EoS parameter $w_{de}$. If the effective sound speed
$c_s\rightarrow 0$, then $V_{de}\rightarrow V_m$, but
$\delta_{de}\rightarrow (1+w_{de})\delta_m$. If we assume that $c_s$ for dark energy is in range (0.1, 1), then in
galaxy for scale range of perturbations from Solar System size to the galaxy one,
$10^{-3}\le\lambda\le10^{4}$ pc, ratio of the amplitudes of velocities is in the range
$10^{-22}-10^{-3}$. In cluster of galaxies for scales from galaxy to cluster of galaxies, $10^{4}\le\lambda\le10^{7}$
pc, ratio of the amplitudes of velocities is in the range $10^{-8}-10^{-2}$. Ratio of amplitudes of density
perturbations is even smaller, since it is multiplied by $(1+w_{de})$, which is in the range (-0.2, 0.2). It
is interesting that in the case of phantom field $\delta_{de}<0$ when $V_{de}<0$. It means that dark energy flows inside
the perturbations (one can determine this process like ``accretion'' in the weak gravitational fields), but the region
of underdensity of dark energy
(void) is created in the places of matter overdensities. Unfortunately, the contrast of density of dark
energy in them is too tiny for search for the observational manifestations. The physical interpretation
of this phenomenon is the same as for increasing of the density of phantom field in the expanding Universe.

In all cases that we analyzed the amplitudes of the oscillation of perturbations of dark energy or their growth in the
potential wells of dark matter density perturbations are many orders of magnitude smaller than the amplitudes of
perturbations of matter on the scales of the considered structures. This conclusion can be also
generalized for the dynamical dark energy with variable $w_{de}$ and $c_s$, what, however, will be the subject of
another paper.

The results obtained here, based on the analytical solutions, are qualitatively consistent with the numerical results
\cite{Dutta2007,Abramo2007,Mota2008,Wang2009,Basse2011,Wang2012} for small perturbations and quantitative differences
are basically caused by differences in features and parameters of different models of dark energy.

In papers \cite{Novosyadlyj2013m,Sergijenko2009b,Novosyadlyj2010,Sergijenko2011,Novosyadlyj2012} we analyzed the
gravitational instability of the matter and dark energy in the expanding Universe and we see that obtained here
analytical solutions and the results of their application to galaxies and galaxy clusters enhance greatly our
understanding of properties of dark energy and its possible manifestations in the Universe.

\section{Conclusions}

The analytical solutions of linear equations for small perturbations of density and velocity of the matter and dark
energy as well as the metric of space-time on the background of static Minkowski world were obtained. The analysis of
solutions was carried out for the parameters of the
models which match well the wide set of cosmological observational data. It is shown that the quintessential dark
energy itself is
gravitationally stable and can only oscillate with constant amplitude on scales smaller than the Jeans scale
$\lambda_J\approx2660(1+3c_s^2)^{-1/2}(1+w_{de})^{-1/2}(6.4\cdot10^{-30}/\rho_{de})^{1/2}$ Mpc, which for realistic
values of $\rho_{de}$ (in g/cm$^3$) is much larger than the scales of observable structures interesting for astrophysics.
Phantom dark energy has no Jeans scale -- the perturbations on all scales can only oscillate with
constant amplitude.

In the two-component medium the amplitudes of the dark matter density perturbations are much larger than dark energy
ones and almost completely determine the gravitational potential of perturbations on given scale. Dark energy can
monotonically flow into gravitational potential wells of positive matter density perturbations ($\delta_m>0$)
oscillating with significantly lower constant amplitude and forming dark energy overdensity
($\delta_{de}>0$) in the case of quintessential dark energy ($-1<w_{de}<-1/3$) and dark energy underdensity
($\delta_{de}<0$) in the case of phantom one ($w_{de}<-1$). We should note,
however, that the amplitudes of perturbations of dark energy at all galaxy and cluster scales are
negligibly small in comparison with amplitudes of perturbations of dark matter ($\delta_{de}\ll\delta_m,\,V_{de}\ll
V_{m}$). Only in the models of dark energy with $c_s\rightarrow 0$ the velocity perturbations of dark
energy are close to velocity perturbations of matter: $V_{de}\rightarrow V_m$, while $\delta_{de}\rightarrow
(1+w_{de})\delta_m$. If
$c_s\ge 0.1$, then $V_{de}\le 10^{-3}V_m$ on the largest scales of galaxies and $V_{de}\le 10^{-2}V_m$ on the largest
scales of clusters of galaxies. So, the amplitudes of perturbations of dark energy may be larger in the dynamical
models of dark energy with small value of speed of sound and may leave traces in the structure on scales of galaxies and
clusters of galaxies.

\section*{Acknowledgements}
This work was supported by the project of Ministry of Education and Science of Ukraine (state registration number
0113U003059) and research program ``Scientific Space Research'' of the National Academy of Sciences of
Ukraine (state registration number 0113U002301).

\end{document}